\documentclass[12pt,english, preprint]{revtex4}
\usepackage[T1]{fontenc}
\usepackage[latin1]{inputenc}
\usepackage{babel}
\usepackage{graphics}

\makeatletter

\providecommand{\LyX}{L\kern-.1667em\lower.25em\hbox{Y}\kern-.125emX\@}

\def\Tr   {\mathop{\hbox{Tr}}}
\def\Real   {\mathop{\hbox{Re}}}
\def\plaq {
\setlength{\unitlength}{1 pt}
\begin{picture}(10,10)(0,0)
\put(0,0){\line(1,0){10}}
\put(0,0){\line(0,1){10}}
\put(10,0){\line(0,1){10}}
\put(0,10){\line(1,0){10}}
\end{picture}
}
\def\dplaq {
\setlength{\unitlength}{1 pt}
\begin{picture}(20,10)(0,0)
\put(0,0){\line(1,0){20}}
\put(0,0){\line(0,1){10}}
\put(20,0){\line(0,1){10}}
\put(10,0){\line(0,1){10}}
\put(0,10){\line(1,0){20}}
\end{picture}
}
\def\chair{
\setlength{\unitlength}{1pt}
\begin{picture}(20,12.33)(0,0)
\put(0,0){\line(1,0){10}}
\put(0,0){\line(3,1){10}}
\put(10,0){\line(0,1){9}}
\put(20,3.33){\line(0,1){9}}
\put(10,3.33){\line(1,0){10}}
\put(10,9){\line(3,1){10}}
\end{picture}
}
\def\skew{
\setlength{\unitlength}{1pt}
\begin{picture}(25,12.33)(0,0)
\put(0,0){\line(1,0){15}}
\put(0,0){\line(3,1){10}}
\put(15,0){\line(0,1){9}}
\put(10,3.33){\line(0,1){9}}
\put(10,12.33){\line(1,0){15}}
\put(15,8.67){\line(3,1){10}}
\end{picture}
}

\makeatother
\begin{document}
{\raggedleft COLO-HEP-481\par}

{\raggedleft March 2002\par}

{\centering \textbf{\Large Evaluating the Fermionic Determinant of
Dynamical Configurations }\Large \par}

\author{Anna Hasenfratz}

\email{anna@eotvos.colorado.edu}

\thanks{}

\affiliation{Department of Physics, University of Colorado, Boulder, CO-80304-0390}

\author{Andrei Alexandru}

\email{alexan@pizero.colorado.edu}

\thanks{}

\affiliation{Department of Physics, University of Colorado, Boulder, CO-80304-0390}

\begin{abstract}
We propose and study  an improved method to calculate the fermionic
determinant of dynamical configurations. The evaluation or at least
stochastic estimation of ratios of fermionic determinants is essential
for a recently proposed updating method of smeared link dynamical
fermions. This update creates a sequence of configurations by changing
a subset of the gauge links by a pure gauge heat bath or over relaxation
step. The acceptance of the proposed configuration depends on the
ratio of the fermionic determinants on the new and original configurations.
We study this ratio as the function of the number of links that are
changed in the heat bath update. We find that even when every link
of a given direction and parity of a 10fm\( ^{4} \) configuration
is updated, the average of the determinant ratio is still close to
one and with the improved stochastic estimator the proposed change
is accepted with about 20\% probability. This improvement suggests
that the new updating technique can be efficient even on large lattices. 

\end{abstract}
\maketitle

\section{introduction}

The use of smeared or fat links in staggered fermion actions has gained
popularity in recent years due to the improved flavor symmetry these
actions possess \cite{Hasenfratz:2001hp,Orginos:1999cr,Lepage:1998vj}.
Smeared links are naturally part of most systematic improvement programs
and many overlap fermion formulations as well \cite{Hasenfratz:2001hr, DeGrand:2000tf}.
The main difficulty that limits the use of smeared link fermions is
their potential complexity in dynamical simulations. Unless the smeared
links are linear combinations of the original thin links the explicit
form of the fermionic force needed for standard molecular dynamics
simulations is very complicated, making the HMC or R algorithms impractical
or even impossible. A recently proposed update for smeared link dynamical
fermions\cite{Knechtli:2000ku, Hasenfratz:2002} avoids this problem
by creating a sequence of configurations by updating a subset of the
gauge links by a pure gauge heat bath or over relaxation step. The
proposed configuration is accepted or rejected according to the change
in the fermionic determinant. In fact one does not even have to evaluate
the change in the determinant, a stochastic estimator can be used
instead. That requires no more than the evaluation of the inverse
fermion matrix on a Gaussian random source vector. 

The above outlined algorithm can fail in two ways. First, if the ratio
of the fermionic determinants is small, the acceptance rate is small.
The algorithm can also fail if the stochastic estimator gives a poor
approximation of the determinant, making the autocorrelation time
of the simulation (i.e. the number of independent Gaussian random
sources needed to get a good estimate for the determinant) very large.
In this paper we discuss systematic ways to improve the stochastic
estimator. With the improved estimator we calculate the ratio of fermionic
determinants as the function of the number of links updated with a
heat bath step, and show that it remains close to one even if the
updated volume is large. We illustrate and test the method using staggered
fermions with HYP smeared gauge links though the generalization to
any other smeared link action is straightforward.

\section{The HYP action and its dynamical update}

In this section we define the HYP action and briefly summarize the
partial-global updating technique. We consider a smeared link action
of the form\begin{equation}
\label{full_action}
S=S_{g}(U)+\bar{S}_{g}(V)+S_{f}(V)
\end{equation}
 where \( S_{g}(U) \) and \( \bar{S}_{g}(V) \) are gauge actions
depending on the thin links \( \{U\} \) and smeared links \( \{V\} \),
respectively, and \( S_{f} \) is the fermionic action depending on
the smeared links only. The updating method and all its improvements
that we discuss in this paper would work with any kind of smeared
links \( \{V\} \), though the efficiency suffers if the smeared links
are not smooth enough. In our work we use HYP smeared links with staggered
fermions. The HYP links are optimized non-perturbatively to be maximally
smooth. The construction and properties of HYP smearing are discussed
in detail in Ref. \cite{Hasenfratz:2001hp}.

We use a plaquette gauge action for \( S_{g}(U) \)\begin{equation}
\label{S_g(U)}
S_{g}(U)=-\frac{\beta }{3}\sum _{p}\Real \Tr (U_{p}).
\end{equation}
 We choose \( \bar{S}_{g}(V) \) to improve computational efficiency
and we will discuss our specific choice in Sect. III.C. The staggered
fermionic matrix is defined in the usual way\begin{equation}
\label{fermion_matrix_M}
M(V)_{i,j}=2m\delta _{ij}+\sum _{\mu }\eta _{i,\mu }(V_{i,\mu }\delta _{i,j-\hat{\mu }}-V_{i-\hat{\mu }\, \mu }^{\dagger }\delta _{i,j+\hat{\mu }}).
\end{equation}
The matrix \( M^{\dagger }(V)M(V) \) is block diagonal on even and
odd lattice sites. In the following we will denote the even block
by \( \Omega  \)\begin{equation}
\Omega (V)=(M^{\dagger }(V)M(V))_{even,even}
\end{equation}
 and define the fermionic action as \begin{equation}
S_{f}(V)=-\frac{n_{f}}{4}\Tr \, \ln \, \Omega (V)
\end{equation}
 to describe \( n_{f} \)  flavors of staggered fermions. In the following
we consider \( n_{f}=4 \) flavors but we will briefly describe the
generalization to arbitrary flavors at the end of Sect. III.B.

In Refs. \cite{Knechtli:2000ku, Hasenfratz:2002} a partial-global
heat bath and over relaxation updating method was proposed to simulate
the system described by Eq. \ref{full_action}. In this paper we are
not concerned about the update itself, but to motivate our interest
in calculating the fermionic determinant ratios we briefly summarize
the main points of the method. In the first step of the update one
changes a subset of the thin links \( \{U\} \) to propose a new thin
gauge link configuration \( \{U'\} \). The new links are chosen with
a heat bath or over relaxed update that satisfies the detailed balance
condition with the thin link gauge action \( S_{g}(U) \). The smeared
links \( \{V\} \) and \( \{V'\} \) are unique once the thin links
are defined. Next the proposed configuration is accepted with the
probability \begin{equation}
\label{Pacc}
P_{\rm {acc}}=\rm {min}\{1,\; \exp (-\bar{S}_{g}(V')+\bar{S}_{g}(V))\frac{\det (\Omega (V'))}{\det (\Omega (V))}\}.
\end{equation}
 The ratio of the determinants can be written as \begin{eqnarray}
\frac{\det (\Omega (V'))}{\det (\Omega (V))}= & \frac{\int d\xi \xi ^{*}\exp (-\xi ^{*}\Omega ^{-1}(V')\Omega (V)\xi )}{\int d\xi \xi ^{*}\exp (-\xi ^{*}\xi )} & \nonumber \\
= & <\exp (-\xi ^{*}[\Omega ^{-1}(V')\Omega (V)-1]\xi )>_{\xi ^{*}\xi } & .\label{det} 
\end{eqnarray}
 The expectation value can be evaluated stochastically where on every
gauge configuration pair \( \{U\} \) and \( \{U'\} \) only one random
source \( \xi  \) is used to estimate the determinant ratio and the
expectation value is taken together with the configuration ensemble
average. That leads to the stochastic acceptance probability \begin{equation}
\label{Pstoch}
P_{\rm {stoch}}=min\{1,e^{-\bar{S}_{g}(V')+\bar{S}_{g}(V)}e^{-\xi ^{*}[\Omega ^{-1}(V')\Omega (V)-1]\xi }\}.
\end{equation}

\section{Improving the partial-global update}

The success of the partial-global updating algorithm depends on two
things. First, on the ratio of the determinants of the new and old
links, and next on the effectiveness of the stochastic estimator.
If the stochastic estimator fluctuates wildly, it can reduce the acceptance
rate to practically zero even if the change in the determinant is
actually small. In the following we will discuss improving the stochastic
estimator first.

\subsection{Improving the stochastic estimator}

To calculate the acceptance probability we have to calculate the ratio
of the determinants 

\begin{equation}
\label{detratio}
\det \, ^{-1}(A)=\frac{\det \Omega '}{\det \Omega }=<\exp (-\xi ^{*}[A-1]\xi )>_{\xi ^{*}\xi }=<\exp (-\Delta S_{f})>
\end{equation}
 where \( \Omega =\Omega (V) \), \( \Omega '=\Omega (V') \) denote
the old and new fermionic matrix, \( A=\Omega '^{-1}\Omega  \) and
\( \xi  \) is a Gaussian random source vector. The standard deviation
of this stochastic estimation can be written as\begin{eqnarray}
\sigma ^{2} & = & <\exp (-2\xi ^{*}[A-1]\xi )>_{\xi ^{*}\xi }-<\exp (-\xi ^{*}[A-1]\xi )>_{\xi ^{*}\xi }^{2}\nonumber \\
 & = & \det \, ^{-1}(2A-1)-\det \, ^{-2}(A).\label{standard_dev} 
\end{eqnarray}
 Eq. \ref{standard_dev} is valid only if the matrix \( 2A-1 \) is
positive definite. If the matrix \( A \) has even one eigenvalue
that is less than or equal to 1/2, the formula in Eq. \ref{standard_dev}
is not valid, the standard deviation is infinite. There is no a priori
reason to assume that the matrix \( A \) has no small eigenvalues.
This is a very serious problem that could make the stochastic estimator
useless in dynamical calculations. If the fermionic matrices \( \Omega  \)
and \( \Omega ' \) are close, i.e. only a few links are changed in
the update, \( A=\Omega '^{-1}\Omega \approx 1 \) and consequently
\( \det (A)\approx 1 \) and \( \det (2A-1)\approx 1 \) as well.
However for an effective updating method we would like to change the
configuration at as many links as possible, which makes the occurrence
of a small eigenvalue likely. In the following we propose a 2-step
solution that can always be used to handle the small eigenvalues in
\( A. \) 

First we follow the program of Refs.\cite{Knechtli:2000ku, Hasenfratz:2002, Hasenbusch:1998yb}
and replace \( \Omega  \) and \( \Omega ' \) by reduced matrices
\begin{equation}
\Omega _{r}=\Omega e^{-2f(\Omega )},\qquad \Omega '_{r}=\Omega 'e^{-2f(\Omega ')}
\end{equation}
 with \( f \) a yet to be determined polynomial . We can rewrite
eq. \ref{detratio} as\begin{eqnarray}
\frac{\det \Omega '}{\det \Omega } & = & \frac{\det \Omega '_{r}}{\det \Omega _{r}}\exp (2\Tr (f(\Omega ')-f(\Omega )))\nonumber \\
 & = & <\exp (-\xi ^{*}[\Omega '^{-1}_{r}\Omega _{r}-1]\xi )>_{\xi ^{*}\xi }\exp (2\Tr (f(\Omega ')-f(\Omega ))).\label{reduced_det_ratio} 
\end{eqnarray}
 Since \( Tr\, f \) can be calculated exactly, only the first factor
of the last expression is evaluated stochastically. Its fluctuations
are minimized if \( A_{r}=\Omega '^{-1}_{r}\Omega _{r}\approx 1 \).
It is difficult to optimize the polynomial \( f \) both for \( \Omega  \)
and \( \Omega ' \) at the same time, instead we choose \( f \) such
that \( \Omega '^{-1}_{r}\approx 1 \). That also guarantees \( \Omega _{r}\approx 1 \).
Since for staggered fermions the eigenvalues of the matrix \( \Omega  \)
can vary between \( 4m^{2} \) and \( 16+4m^{2} \), we choose the
polynomial \( f \) such the the function \( e^{2f(x)}/x \) is close
to one in that range. In practice we use a third order polynomial
\begin{equation}
\label{def_f}
f(x)=\alpha _{0}+\alpha _{2}x+\alpha _{4}x^{2}+\alpha _{6}x^{3}
\end{equation}
and choose the coefficients \( \alpha _{i} \) by minimizing the function
\begin{equation}
\label{min for f}
\Delta =\int ^{16+4m^{2}}_{4m^{2}}(\frac{1}{x}e^{2f(x)}-1)^{2}\rho (x)dx.
\end{equation}
 The weight function \( \rho (x) \) should approximate the eigenvalue
density distribution of the fermionic matrix. We used a linear approximation
for \( \rho  \) \begin{eqnarray}
\rho (x) & = & x,\qquad \qquad \qquad \qquad x\in (4m^{2},8+4m^{2})\nonumber \\
 & = & 16+8m^{2}-x,\qquad \; \; x\in (8+4m^{2},16+4m^{2})\label{weight_fn} 
\end{eqnarray}
and considered mass values \( m=0.01-0.1 \). We have also tried more
complex forms for the eigenvalue density that included higher order
terms, all motivated by free field calculations. The results were
not very sensitive to the specific choice of \( \rho  \). We do not
want to change the \( \alpha  \) parameters depending on the quark
mass of the simulation so we decided to use the following values in
all cases\begin{eqnarray}
\alpha _{0} & = & -0.34017\nonumber \\
\alpha _{2} & = & \; 0.35645\nonumber \\
\alpha _{4} & = & -0.030379\nonumber \\
\alpha _{6} & = & \; 0.000957.\label{alpha_param} 
\end{eqnarray}

The eigenvalues of the reduced matrix \( \Omega _{r} \) span a smaller
range than the original fermionic matrix. At mass \( am=0.1 \) the
smallest eigenvalue is increased from \( 4m^{2}=0.04 \) to about
\( 0.08 \), while the ratio of the largest to smallest eigenvalue
is reduced from \( (16+4m^{2})/4m^{2}\approx 400 \) to about 14.
At \( am=0.04 \) the increase in the smallest eigenvalue is from
\( 0.0064 \) to \( 0.0125 \), while the reduction in the ratio of
the largest to smallest eigenvalue is from 2,500 to about 95.

When expressed in terms of the reduced matrices the acceptance probabilities
of Eqs. \ref{Pacc},\ref{Pstoch} contain a new gauge-action like
term \begin{eqnarray}
P_{\rm {acc}} & = & \rm {min}\{1,\; \exp (-\Delta \bar{S}_{g}+2\Delta f)<\exp (-\xi ^{*}[\Omega '^{-1}_{r}\Omega _{r}-1]\xi )>_{\xi ^{*}\xi }\}\nonumber \\
P_{\rm {stoch}} & = & \rm {min}\{1,\; \exp (-\Delta \bar{S}_{g}+2\Delta f)\exp (-\xi ^{*}[\Omega '^{-1}_{r}\Omega _{r}-1]\xi )\}\label{Pacc_reduced} 
\end{eqnarray}
 with \( \Delta \bar{S}_{g}=\bar{S}_{g}(V')-\bar{S}_{g}(V) \) and
\( \Delta f=\Tr (f(\Omega ')-f(\Omega )) \). To calculate the expression
\( \xi ^{*}[\Omega '^{-1}_{r}\Omega _{r}-1]\xi  \)  requires multiplications
with \( \Omega  \), \( \Omega '^{-1} \), and with the reduction
factors \( \exp (-2f(\Omega )) \) and \( \exp (2f(\Omega ')) \).
The exponentials can be expanded in a Taylor series and approximated
with a few terms. In \cite{Hasenfratz:2002} we found that it was
sufficient to keep only 15 terms. Later we will argue that it is better
to replace \( \Omega _{r} \) and \( \Omega _{r}'^{-1} \) themselves
with a finite order polynomial. 

To complete the evaluation of the determinant and acceptance probabilities
we still have to calculate the trace of \( f(\Omega ) \). \( \Tr f(\Omega ) \)
can be expressed as the combination of the plaquette and the three
6-link loops of the smeared links \( V. \) It is a fairly straightforward
calculation giving\begin{eqnarray}
\Tr f(\Omega ) & = & (-8\beta _{4}+336\beta _{6})\sum _{n}\Real \Tr \plaq _{n}+\nonumber \\
 &  & 12\beta _{6}[-\sum _{n}\Real \Tr \dplaq _{n}-\sum _{n}\Real \Tr \chair _{n}+\nonumber \label{f_explicit} \\
 &  & \sum _{n}\Real \Tr \skew _{n}-\delta _{N_{t},6}\sum _{n}\Real \Tr P_{n}]+\rm {const},\label{f_explicit} 
\end{eqnarray}
where the summation is over all distinct objects in the lattice. \( P_{n} \)
is the Polyakov line that gives a contribution on lattices of size
\( N=6 \). Lattices that are smaller than 5 in any direction would
have additional contribution of length-6 overlapping loops. Eq. \ref{f_explicit}
is not valid in that case. The coefficients \( \beta  \) are related
to the quark mass and the optimized \( \alpha  \) parameters of Eq.
\ref{alpha_param} as \begin{eqnarray}
\beta _{0} & = & \alpha _{0}+\alpha _{2}\cdot 4m^{2}+\alpha _{4}\cdot (4m^{2})^{2}+\alpha _{6}\cdot (4m^{2})^{3}\nonumber \\
\beta _{2} & = & -\alpha _{2}-\alpha _{4}\cdot 2\cdot 4m^{2}-\alpha _{6}\cdot 3\cdot (4m^{2})^{2}\nonumber \\
\beta _{4} & = & \alpha _{4}+\alpha _{6}\cdot 3\cdot 4m^{2}\nonumber \\
\beta _{6} & = & -\alpha _{6}.
\end{eqnarray}

In Ref. \cite{Knechtli:2000ku} we used a similar reduction using
a second order polynomial for \( f. \) There we determined the \( \alpha _{2} \)
and \( \alpha _{4} \) coefficients by trial and error attempting
to maximize the acceptance rate in Eq. \ref{Pacc}. The values we
obtained there are consistent with what we would determine now with
the minimization procedure. Increasing the order of the reduction
polynomial further gives only slight improvement but would require
the evaluation of \( \Tr \, \Omega ^{4} \) and higher order terms
in \( \Tr \, f \), a considerable computational task.

The polynomial reduction of the fermionic matrix results in considerable
improvement in the evaluation of the determinant ratio but it is not
sufficient to guarantee that the standard deviation of the stochastic
estimator is finite or small. To achieve that we now proceed to rewrite
the reduced fermionic determinant ratio as\begin{eqnarray}
\det \, ^{-1}A_{r} & = & \det \, ^{-n}A_{r}^{1/n}\nonumber \\
 & = & <\rm {exp}(-\sum ^{n}_{j=1}\xi ^{*}_{j}[A_{r}^{1/n}-1]\xi _{j})>_{\xi _{j}^{*}\xi _{j}}\label{det_1/n} 
\end{eqnarray}
with n an arbitrary positive integer. The expectation value is evaluated
with n independent \( \xi _{j} \) random vectors and the standard
deviation becomes \begin{equation}
\label{sigma2_1/n}
\sigma ^{2}=\det \, ^{-n}(2A_{r}^{1/n}-1)-\det \, ^{-2}(A_{r}).
\end{equation}
 \( \sigma ^{2} \) is finite if none of the eigenvalues of the matrix
\( A_{r} \) is smaller than or equal to \( 2^{-n} \). This is a
much easier condition to satisfy then the one before. With the reduced
matrix, assuming that the smallest eigenvalue of \( A_{r} \) is not
smaller than the product of the smallest eigenvalues of \( \Omega _{r} \)
and \( \Omega _{r}'^{-1} \), \( n=4 \) is sufficient to guarantee
the finiteness of the standard deviation at \( am=0.1, \) and \( n=8 \)
is sufficient at \( am=0.04 \). For small standard deviation one
might need to use larger \( n \) values, but with any mass \( am\neq 0 \)
and matrix \( A_{r} \) it is possible to choose \( n \) such that
the standard deviation is finite and small. The cost of this improvement
is that we have to evaluate the expression \( \xi ^{*}(A^{1/n}_{r}-1)\xi  \)
n times for one estimate of the determinant. 

The nth root of the matrix \( A_{r} \) can be approximated with polynomials
to arbitrary precision \cite{Montvay:1996ea, Montvay:1997vh}. Since
the order of the matrices in the determinant are irrelevant, we write
the nth root of the matrix as \begin{equation}
\label{A_1/n}
A_{r}^{1/n}=\Omega '\, _{r}^{-1/2n}\, \Omega ^{1/n}_{r}\, \Omega '\, _{r}^{-1/2n}.
\end{equation}
 We found that breaking up \( \Omega '\, _{r}^{-1/n} \) to two identical
terms and separating them with the less singular \( \Omega ^{1/n}_{r} \)
improves the stochastic estimator. The terms \( \Omega ^{1/n}_{r} \)
and \( \Omega '\, _{r}^{-1/2n} \) are approximated with polynomials
\begin{eqnarray}
\Omega '\, ^{-1/2n}_{r} & =\Omega '\, ^{-1/2n}\exp (f(\Omega ')/n)= & P_{l}^{(2n)}(\Omega '),\nonumber \\
\Omega _{r}^{1/n} & =\Omega ^{1/n}\exp (-2f(\Omega )/n)= & Q^{(n)}_{k}(\Omega ),\label{poly_forms} 
\end{eqnarray}
 where \( P^{(2n)}_{l} \) and \( Q^{(n)}_{k} \) are \( l \) and
\( k \) order polynomials of the fermionic matrices \( \Omega  \)
and \( \Omega ' \). To reduce the errors of the polynomial approximation
we write the exponent in the stochastic estimator as \begin{equation}
\label{stoch_poly}
\xi ^{*}[A_{r}^{1/n}-1]\xi =\xi ^{*}P_{l}^{(2n)}(\Omega ')\, Q_{k}^{(n)}(\Omega )P_{l}^{(2n)}\, (\Omega ')\xi -\xi ^{*}P_{l}^{(2n)}(\Omega ')\, Q_{k}^{(n)}(\Omega ')\, P_{l}^{(2n)}(\Omega ')\xi .
\end{equation}
 The necessary order for the polynomials \( P \) and Q vary with
the quark mass but we found that in most cases fairly low orders are
sufficient. 

At this point the generalization of the partial-global updating method
to arbitrary flavor number is straightforward. To describe \( n_{f} \)
flavors we have to replace the determinant ratio in Eq. \ref{Pacc}
by its \( n_{f}/4 \)th power. That can be easily done by summing
up to \( \frac{n_{f}}{4}n \) only in Eq. \ref{det_1/n}. The polynomials
\( P \) and Q do not have to be changed and smaller \( n \) will
be sufficient for the same standard deviation of the determinant.

\subsection{Calculating the determinant ratio}

To illustrate the improved estimator, in this section we calculate
the ratio \( \det (\Omega ')/\det (\Omega ) \) for a specific configuration
pair. We chose \( \{U\} \) from a configuration set that was generated
with the HYP dynamical action at \( \beta =5.2 \), \( am=0.1 \),
and \( \bar{S}_{g}(V)=0 \). The scale at these parameter values is
\( r_{0}/a=3.0(1) \) and \( m_{\pi }/m_{\rho }\approx 0.8 \) \cite{Hasenfratz:2002}.
We updated 300 random links of \( \{U\} \) with a heat bath step
corresponding to a \( \beta =5.2 \) plaquette gauge action to create
the \( \{U'\} \) configuration. To calculate the determinant ratio
we use Eq. \ref{reduced_det_ratio} with \( f=0 \), with the improved
\( f \) given in Eqs. \ref{def_f}, \ref{alpha_param}, and also
using the formula of Eq. \ref{det_1/n} with \( n=8 \) and the improved
\( f. \) We calculate the expectation value using 500-1500 random
vectors. 
\begin{figure}
{\centering \resizebox*{13cm}{!}{\includegraphics{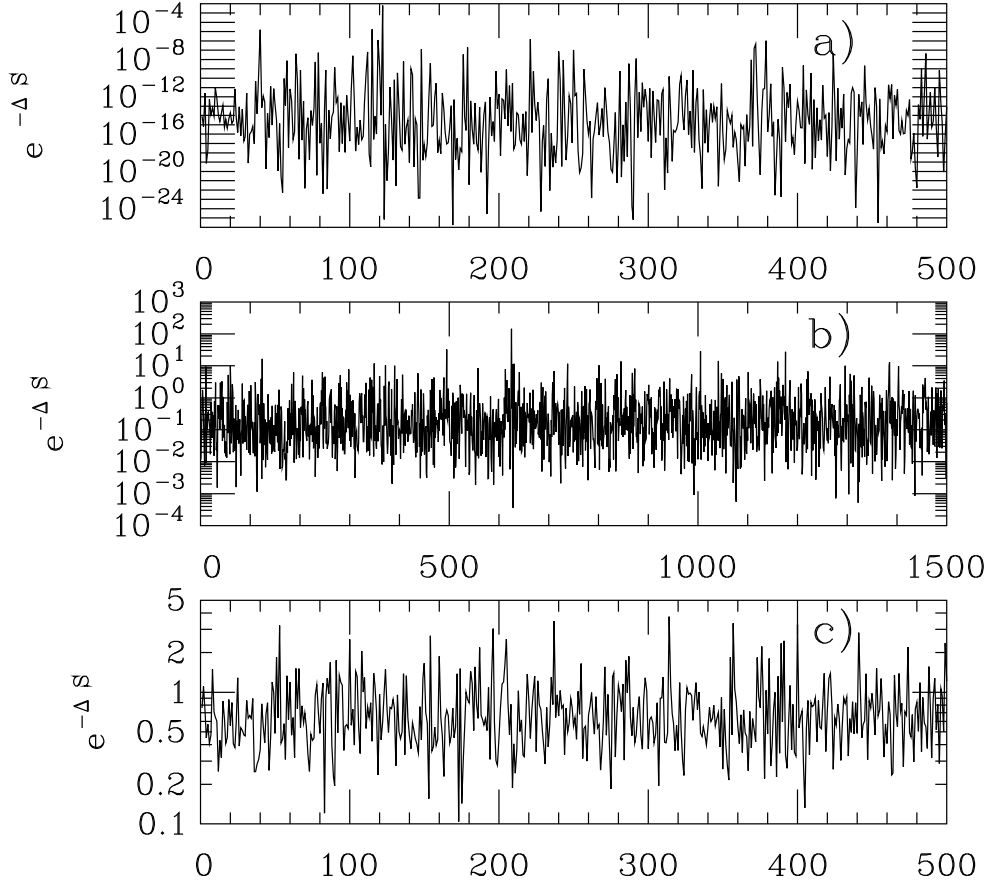}} \par}

\caption{The stochastic estimator for \protect\( \det (\Omega ')/\det (\Omega )\protect \)
using a) the naive estimator with \protect\( f=0\protect \) , b)
the estimator with improved \protect\( f\protect \) given in Eqs.
\ref{def_f}, \ref{alpha_param} and c) the form of Eq. \ref{det_1/n}
with \protect\( n=8\protect \) and improved \protect\( f\protect \)
.\label{detb5.2} }
\end{figure}
 Figure \ref{detb5.2} shows the stochastic estimator for the three
cases. One could not guess from the figure that the three estimators
describe the same quantity. The naive \( f=0 \) estimator is 15 orders
of magnitude smaller than the improved ones. How is that possible?
The average \( <e^{-\Delta S_{f}}> \) will be the same for all three
estimators, but for the naive one the average will come from many
almost zero values and an occasional large one. That occasional large
value is so rare that we did not even encounter it in 500 samples.
The improved estimator with \( n=1 \) looks much more reliable and
it predicts \begin{equation}
\frac{\det (\Omega ')}{\det (\Omega )}=0.81(12),\qquad n=1.
\end{equation}
 The estimator with \( n=8 \) is even better. With only a third of
the statistics of the \( n=1 \) case it predicts the determinant
ratio as \begin{equation}
\frac{\det (\Omega ')}{\det (\Omega )}=0.77(2),\qquad n=8.
\end{equation}
 That does not mean that the acceptance rate of the partial-global
update is close to 80\% if we update 300 links at a time. The acceptance
rate is better described by the expectation value \begin{eqnarray}
<\rm {min}\{1,\; e^{-\Delta S_{f}}\}>_{\xi ^{*}\xi }= & 0.32(1),\qquad  & n=1,\nonumber \\
= & 0.65(1),\qquad  & n=8.
\end{eqnarray}
 One should remember that the above values correspond to a specific
pair of configurations. Before calculating the determinant ratio on
an ensemble of configurations we first discuss a modification of the
gauge action.

\subsection{Choosing the gauge action \protect\( \bar{S}_{g}(V)\protect \) }

In the previous chapter we showed how to remove the most singular
part of the inverse fermion matrix by multiplying it with a factor
\( \exp (2f(\Omega )) \). The change in the fermion determinant is
compensated by the additional term \( \exp (2\Tr (f(\Omega ')-f(\Omega ))) \)
in the stochastic estimator that can be calculated exactly. With \( f(\Omega ) \)
a third order polynomial \( \Tr \, f(\Omega ) \) is a combination
of the plaquette and 6-link loops of the smeared links as given in
Eq. \ref{f_explicit}. While it is straightforward to evaluate \( f(\Omega ) \),
it is not a completely negligible computational cost. On the smeared
link lattice the plaquette and the 6-link loops are very correlated
and \( \Tr \, f(\Omega ) \) can be approximated by the plaquette
term only, thus reducing the computational overhead. In general we
choose the gauge action \( \bar{S}_{g}(V) \) as \begin{equation}
\label{S_g}
\bar{S}_{g}(V)=2\Tr \, f(\Omega )-12\beta _{6}\delta _{N_{t},6}\sum _{n}\Real \Tr P_{n}-\frac{\gamma }{3}\sum _{n}\Real \Tr \plaq _{n}.
\end{equation}
 Here we have included the Polyakov line term explicitly, i.e. \( \bar{S}_{g}(V) \)
is the combination of the 4 and 6-link gauge loops only, it contains
no loops closed because of the periodicity of the lattice. Like before,
this action should not be used if in any direction the lattice is
smaller than 5. With this choice for \( \bar{S}_{g}(V) \) the gauge
term in the acceptance probability in Eq. \ref{Pacc_reduced} simplifies
to \begin{equation}
-\Delta \bar{S}_{g}+2\Delta f=12\beta _{6}\delta _{N_{t},6}\sum _{n}(\Real \Tr P_{n}'-\Real \Tr P_{n})+\frac{\gamma }{3}\sum _{n}(\Real \Tr \plaq _{n}-\Real \Tr \plaq _{n}\, ').
\end{equation}
 We can choose the coefficient \( \gamma  \) to account for the \( 2\Tr \, f(\Omega ) \)
term, or even better, we can choose it to maximize the determinant
ratios and the acceptance rate. Then we not only avoid the computation
of the term \( \exp (2\Tr (f(\Omega ')-f(\Omega ))) \) in Eq. \ref{Pacc_reduced}
but can also increase the efficiency of the updating algorithm. 

When we write the gauge action as \( S_{g}(U)+\bar{S}_{g}(V) \),
we break up the gauge term into two pieces. We use the first term
\( S_{g}(U) \) in the heat bath update and include the second one,
\( \bar{S}_{g}(V) \), in the accept-reject term. Such a break-up
usually lowers the acceptance rate, especially if the second term
fluctuates considerably. With the choice of Eq. \ref{S_g} \( \bar{S}_{g}(V) \)
actually cancels an other fluctuating term, \( 2\Delta f \), and
the algorithm should get more efficient. The introduction of the plaquette
term proportional to \( \gamma  \) could compromise the efficiency.
Since the smeared plaquette term does not fluctuate much, a small
\( \gamma  \) coefficient does not harm the acceptance rate much.
How should we choose the coefficient \( \gamma  \)? According to
Ref. \cite{Hasenfratz:2002} the HYP action with \( \bar{S}_{g}(V)=0 \)
at \( \beta =5.2 \), \( am=0.1 \) has lattice spacing \( a\approx 0.17 \)fm.
In the global heat bath update the links of the configurations are
updated with a pure gauge action of gauge coupling \( \beta =5.2 \).
The pure gauge configurations at this coupling are very different
form the dynamical configurations. The correlation length that characterizes
the large distance behavior is much smaller on the pure gauge configurations.
At short distances the average plaquette on the dynamical configurations
is \( <\Real \Tr \plaq >_{dyn}=1.45 \) while on the pure gauge configurations
the average plaquette is much smaller, \( <\Real \Tr \plaq >_{\beta =5.2}=1.30 \).
The gauge action that we use to create new configurations does not
match the dynamical action neither at long nor at short distances.
One would expect that the partial-global update is most effective
when the pure gauge configurations of the heat bath step are close
to the dynamical configurations. This suggests that in order to maximize
the efficiency of the partial-global update we can try to match the
short and/or long distance fluctuations of the heat bath and dynamical
actions. To match the short distance fluctuations we can require that
the average plaquette of the heat bath update action and the dynamical
action are close. This condition requires different \( \gamma  \)
coupling at different quark masses and gauge coupling values but choosing
\begin{equation}
\label{gamma}
\gamma =-0.1
\end{equation}
 offers a good compromise. The choice \( \gamma =0.0 \) is not much
worse and leads to a somewhat simpler action, but in the following
we will use \( \gamma =-0.1 \). By construction now the small scale
fluctuations of the pure gauge heat bath action and of the dynamical
action are about the same. The modification also improves the matching
of the large distance correlations. With the new action, in order
to reproduce the lattice spacing \( a=0.17 \)fm of the \( \beta =5.2, \)
\( am=0.1 \), \( \bar{S}_{g}(V)=0 \) action, we have to choose the
gauge coupling \( \beta =5.65 \) while keeping \( am=0.1 \). The
lattice spacing of the pure gauge model at \( \beta =5.65 \) is almost
\( 0.17 \)fm, a very good agreement. At smaller quark masses the
same lattice spacing requires a smaller gauge coupling \( \beta  \)
suggesting that a slightly larger \( \gamma  \) value would be the
optimal one. At this point we feel that the difference is not significant
to justify a quark mass dependent coupling. 

By choosing the gauge action \( \bar{S}_{g}(V) \) according to Eq.
\ref{S_g} we modify the dynamical action. This modification will
not affect the perturbative properties of the smeared link action
as the terms in \( \bar{S}_{g}(V) \) are independent of the thin
link gauge coupling and will become negligible in the continuum limit.
At finite lattice spacing the new terms could change the scaling behavior
of the system and their effect should be investigated.

\section{Determinant ratios with the modified action}

In this chapter we investigate the fermionic determinant ratios on
configurations generated with the modified action of Eqs. \ref{S_g},
\ref{gamma}. We will use two sets of configurations. Both sets contain
about 100 \( 8^{3}24 \) lattices. The first one was created at \( \beta =5.65 \),
\( am=0.1 \) and has lattice spacing \( a=0.17 \)fm (\( r_{0}/a=2.95(5) \))
and \( m_{\pi }/m_{\rho }\approx 0.70 \). The second set is at couplings
\( \beta =5.55 \), \( am=0.04 \) with lattice spacing \( a=0.17 \)fm
(\( r_{0}/a=2.88(6) \)) and \( m_{\pi }/m_{\rho }\approx 0.55 \).
On both sets we created pairs of configurations by updating a random
subset of the original thin links with a heat bath step corresponding
to the thin link pure gauge couplings, i.e. \( \beta =5.65 \) and
\( \beta =5.55 \). On each pair we calculated the modified determinant
ratio \( \exp (-\Delta \bar{S}_{g})\det ^{-1}(\Omega \Omega '^{-1}) \)
using Eqs. \ref{det},\ref{det_1/n},\ref{stoch_poly} with 400(800)
random source vectors with \( n=4 \)(8) break-up of the determinant,
i.e. we estimated the determinant value from 100 independent measurements
on each configuration pair. 
\begin{figure}
{\centering \resizebox*{17cm}{!}{\includegraphics{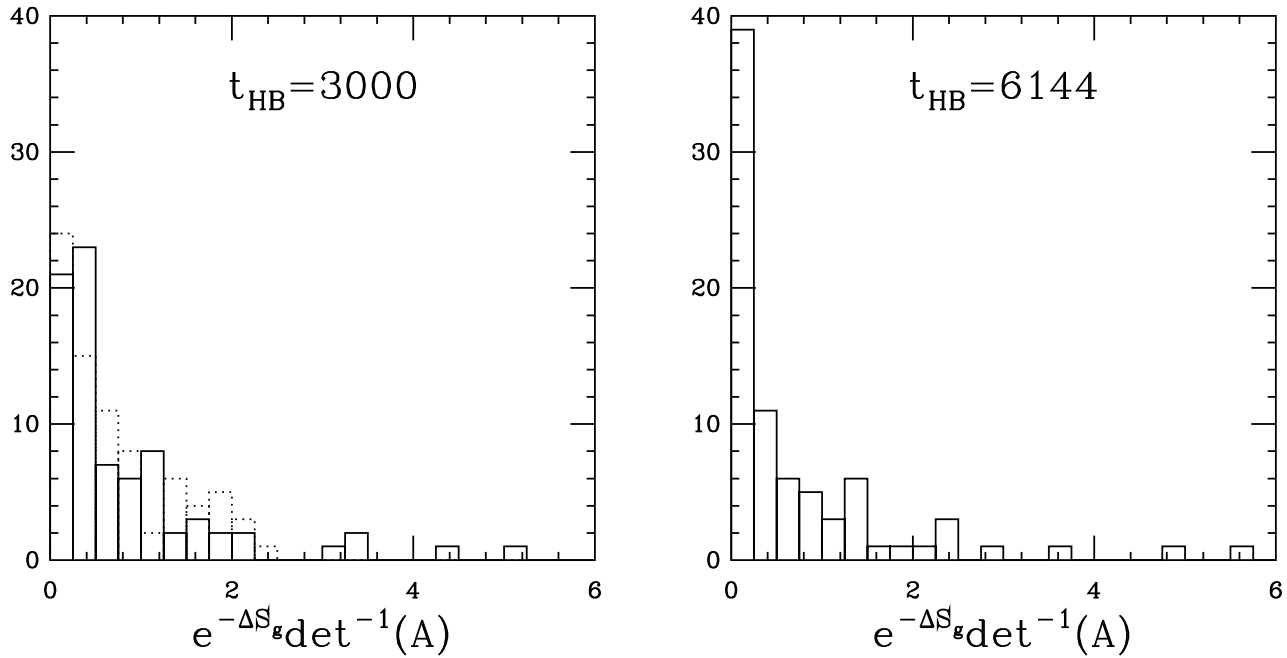}} \par}

\caption{The distribution of the modified fermionic determinant ratios on
configuration set I. a) \protect\( t_{HB}=3000\protect \) links are
updated with a heat bath step and the determinant ratios are calculated
with \protect\( n=4\protect \) (dotted lines) and \protect\( n=8\protect \)
(solid lines) determinant break-up. b) all links of a given direction
and parity (\protect\( t_{HB}=6144)\protect \) are updated at once
and the determinant ratio is calculated with \protect\( n=8\protect \)
break-up.\label{plot_det_5.65} }
\end{figure}
We calculated the determinant both with relatively small order polynomials
(order 16 to 32) and higher order polynomials (order 64 to 128) to
monitor possible systematical errors. The difference between the small
and higher order approximations is small and well within the errors
of the final results. The numbers we present here were obtained with
the higher order polynomials. 

Figure \ref{plot_det_5.65} shows the distribution of the modified
determinant ratios on 80 configuration pairs from set I. The histogram
of figure \ref{plot_det_5.65}/a corresponds to determinant ratios
on configuration pairs that differ at 3000 links. The solid lines
shows the distribution measured with \( n=8 \), the dotted lines
with \( n=4 \) determinant break-up. The two measurements are consistent
predicting \begin{equation}
<e^{-\Delta \bar{S}_{g}}\det \, ^{-1}(\Omega \Omega '^{-1})>_{\rm {set\, I}}=0.80(13),\qquad t_{HB}=3000
\end{equation}
 for the determinant and \begin{equation}
<\rm {min}\{1,e^{-\Delta \bar{S}_{g}}\det \, ^{-1}(\Omega \Omega '^{-1})\}>_{\rm {set\, I}}=0.54(5),\qquad t_{HB}=3000
\end{equation}
 for the acceptance rate as defined in Eq. \ref{Pacc}. While the
\( n=4 \) and \( n=8 \) measurements agree in their prediction of
the determinant ratios, they differ considerably in their standard
deviation. The average standard deviation as defined in Eq. \ref{sigma2_1/n}
of the \( n=4 \) calculation is \( \sigma _{n=4}=3.5(8) \) while
for \( n=8 \) it is \( \sigma _{n=8}=1.8(3) \). The standard deviation
of the determinant measurement can influence the autocorrelation time
of a simulation as that depends both on the effectiveness of the gauge
update and on the error of the stochastic estimator. A factor of two
increase in the standard deviation of the determinant could require
up to a factor of four increase in the number of stochastic estimators,
increasing the autocorrelation time accordingly. The extra computational
cost of breaking the determinant up to \( n=8 \) instead of \( n=4 \)
parts could be easily compensated with the reduced autocorrelation
time. Whether it is worth using even larger number of terms should
be investigated at different quark masses separately. 
\begin{figure}
{\centering \resizebox*{9cm}{!}{\includegraphics{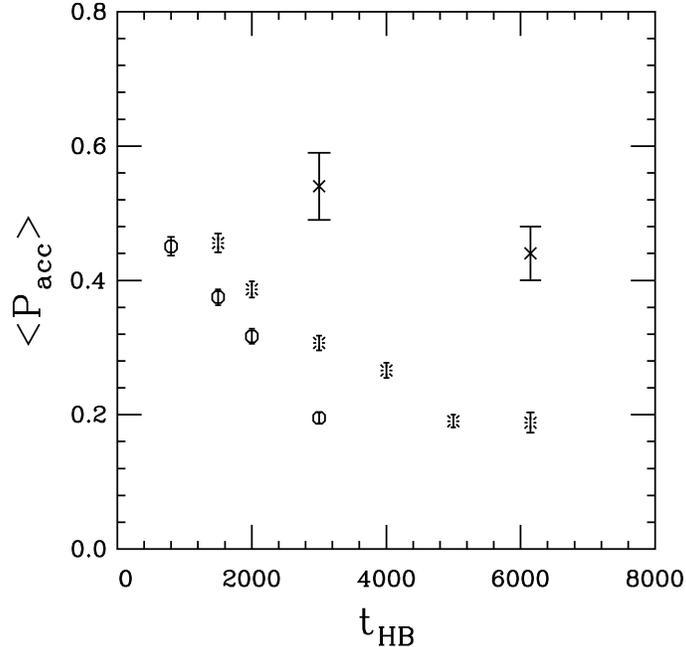}} \par}

\caption{The configuration average of the stochastic acceptance rate \protect\( <P_{\rm {stoch}}>_{U}\protect \)
on the configuration set I as the function of the number of links
touched in the heat bath update. Bursts correspond to \protect\( n=8\protect \),
octagons to \protect\( n=4\protect \) break-up of the determinant.
The crosses correspond to the acceptance rate using the determinants
as defined in Eq. \ref{Pacc}.\label{Pacc_5.65}}
\end{figure}

With the heat bath update we change a random set of links of given
direction and parity. On an \( 8^{3}24 \) configuration a maximum
of 6144 links can be changed at once. In figure \ref{plot_det_5.65}/b
we show the modified determinant ratio distribution when we update
all 6144 links of a randomly chosen direction and parity. This result
was obtained with \( n=8 \) break-up. The average of the determinant
ratios \begin{equation}
<e^{-\Delta \bar{S}_{g}}\det \, ^{-1}(\Omega \Omega '^{-1})>_{\rm {set\, I}}=0.71(10),\qquad t_{HB}=6144,
\end{equation}
 and the acceptance rate as defined in Eq. \ref{Pacc} \begin{equation}
<\rm {min}\{1,e^{-\Delta \bar{S}_{g}}\det \, ^{-1}(\Omega \Omega '^{-1})\}>_{\rm {set\, I}}=0.44(4),\qquad t_{HB}=6144
\end{equation}
 are not much different from the previous \( t_{HB}=3000 \) values,
though the average standard deviation is worse, \( \sigma _{n=8}=2.7(6) \).
To us this is a surprising result: on a 10fm\( ^{4} \), \( 8^{3}24 \)
lattice we can perform a heat bath update on all the links in a given
direction and parity, the maximum that can be updated on this volume
simultaneously, and accept this change with close to 50\% probability.
The configuration average of the stochastic acceptance rate \( <P_{\rm {stoch}}>_{U} \)
of Eq. \ref{Pstoch} is not that high. Figure \ref{Pacc_5.65} compares
the average stochastic acceptance rate as the function of the links
touched in the heat bath update both for \( n=4 \) and \( n=8 \)
determinant break-up and the acceptance rate from the determinant
as defined in Eq. \ref{Pacc}. With \( n=8 \) the stochastic acceptance
rate is close to 20\% if \( t_{HB}=6144 \) and about 30\% if \( t_{HB}=3000 \).
The stochastic acceptance rate with \( n=4 \) is somewhat lower.
Even though these values are smaller than the maximal ones predicted
by the determinants themselves, they are still quite large. What parameters
would provide the best choice in an actual simulation depends on many
things: the number of links that effectively change in an update step,
the cost of increasing the breakup of the determinant, and on the
autocorrelation time of the simulation. The study of these questions
is beyond the scope of the present paper and we will return to them
in a future publication.
\begin{figure}
{\centering \resizebox*{8cm}{!}{\includegraphics{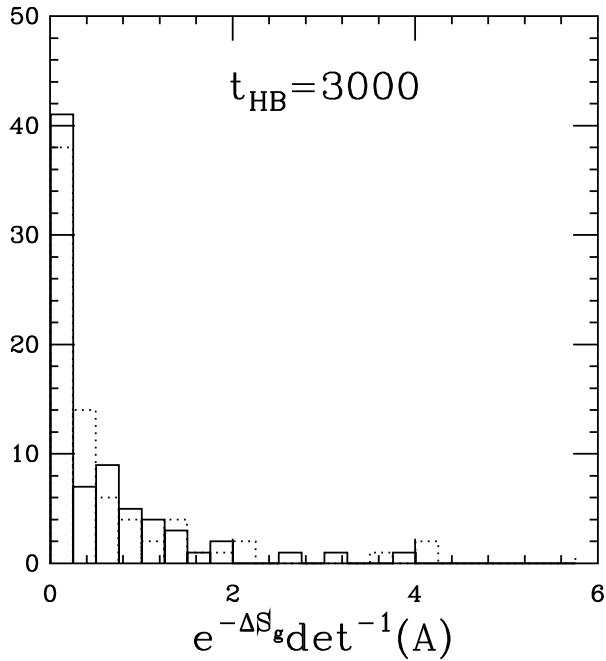}} \par}

\caption{The distribution of the modified fermionic determinant ratio on configuration
set II. \protect\( t_{HB}=3000\protect \) links are updated with
a heat bath step and the determinant ratios are calculated with \protect\( n=4\protect \)
(dotted lines) and \protect\( n=8\protect \) (solid lines) break-up.
\label{plot_det_5.55}}
\end{figure}

\begin{figure}
{\centering \resizebox*{9cm}{!}{\includegraphics{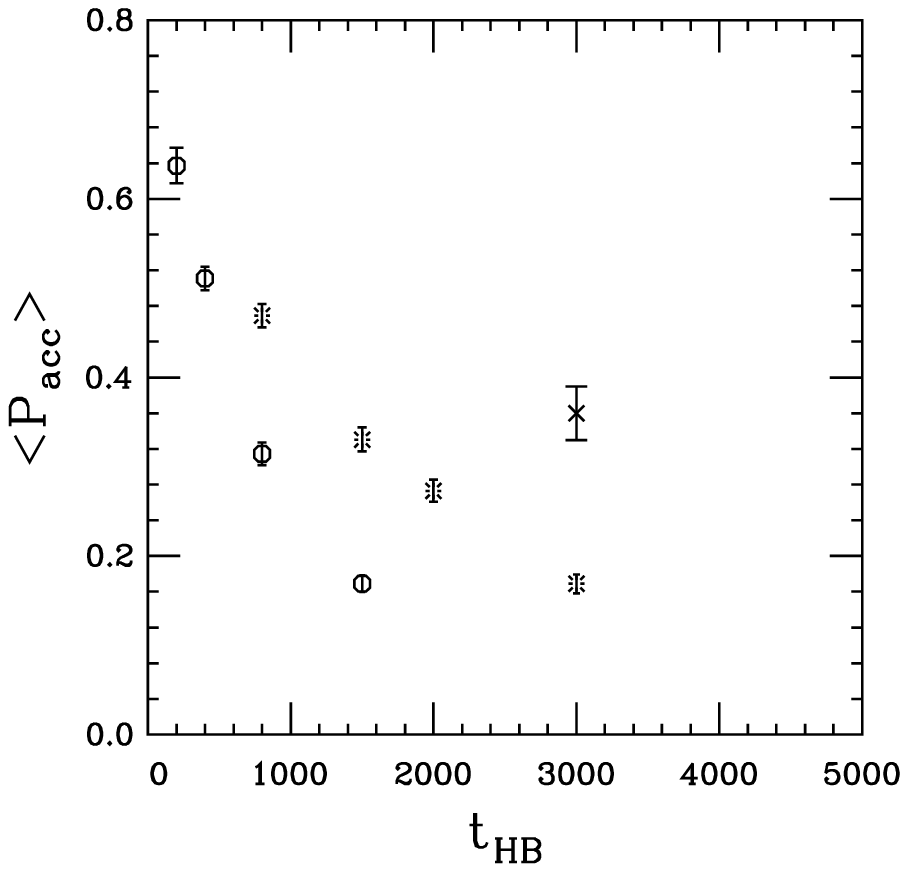}}  \par}

\caption{The configuration average of the stochastic acceptance rate \protect\( <P_{\rm {stoch}}>_{U}\protect \)
on the configuration set II as the function of the number of links
touched in the heat bath update. Bursts correspond to \protect\( n=8\protect \),
octagons to \protect\( n=4\protect \) break-up of the determinant.
The cross corresponds to the acceptance rate using the determinants
as defined in Eq. \ref{Pacc}.\label{Pacc_5.55}}
\end{figure}

Figures \ref{plot_det_5.55} and \ref{Pacc_5.55} are the same as
figures \ref{plot_det_5.65}/a and \ref{Pacc_5.65} but for configuration
set II. Most fermionic methods lose some efficiency at smaller quark
masses and the stochastic estimator is no exception. While the average
of the determinant ratios \begin{equation}
<e^{-\Delta \bar{S}_{g}}\det \, ^{-1}(\Omega \Omega '^{-1})>_{\rm {set\, II}}=0.8(2),\qquad t_{HB}=3000,
\end{equation}
 is not much different from the set I configurations, the acceptance
rate as defined in Eq. \ref{Pacc} is smaller \begin{equation}
<\rm {min}\{1,e^{-\Delta \bar{S}_{g}}\det \, ^{-1}(\Omega \Omega '^{-1})\}>_{\rm {set\, II}}=0.35(7),\qquad t_{HB}=3000.
\end{equation}
 Whether this decrease is due to the smaller quark mass or reflects
the fact that the pure gauge heat bath action does not match the dynamical
action well is worth further investigation. To match the stochastic
acceptance rate of set I with \( n=4 \) determinant break-up we have
to use \( n=8 \) on set II as figure \ref{Pacc_5.55} shows.

\section{Conclusion}

In this paper we proposed an improved method to calculate the fermionic
determinant of dynamical configurations. The method is very general
but relies on the smoothness of smeared gauge links. To test the method
we considered dynamical configurations, updated a large subset of
their links with a pure gauge heat bath step, and calculated the ratios
of the fermionic determinants on the old and new configurations. We
found that even if all the links of a given direction and parity of
an \( 8^{3}24 \), 10fm\( ^{4} \), \( m_{\rho }/m_{\pi }=0.7 \)
configuration are updated at once, (6144 links in all), the fermionic
determinant ratio is still fairly large and such a change would be
accepted by a Metropolis accept-reject step with about 50\% probability.
Using only a single stochastic estimator for the determinant reduces
the acceptance rate to 20\% but still offers an effective update.
On configurations with smaller quark masses the stochastic estimator
loses some efficiency. When \( m_{\rho }/m_{\pi }=0.55 \) only about
half that many links can be updated at one time with 20\% stochastic
acceptance rate though the determinants stay about the same as with
larger quark masses. 

We have not used the fully improved method in dynamical simulations
yet, nor did we optimize all its parameters. The optimization requires
tuning the parameters of the action and calculating autocorrelation
times with different determinant break-up and updating steps. This
work is in progress and the results will be reported in a forthcoming
publication.

\begin{acknowledgments}
A. H. would like to thank the hospitality of the Institute of Nuclear
Theory, University of Washington at Seattle, where part of this work
was carried out. She is particularly indebted to Prof. I. Montvay
who not only shared his polynomial fitting code but gave valuable
help and advice in modifying it for this project. We received useful
comments regarding this manuscript from Profs. T. DeGrand and P. Hasenfratz.
A.H. benefited from many inspiring conversations with Francesco Knechtli. 
\end{acknowledgments}
\bibliographystyle{apsrev}
\bibliography{lattice}

\end{document}